\documentclass[reprint,amsmath,amssymb,aps,superscriptaddress]{revtex4}

\usepackage{xr}
\usepackage{graphicx}
\usepackage{epstopdf}

\begin{document}

\title{A corner reflector of graphene Dirac fermions as a phonon-scattering sensor}
% Total word count 5000 words (excluding abstract, methods, references (<70) and figure legends ((<350 words), footnotes? (no footnotes)). Title 15 words. Abstract 150 words. Intro sans titre. Sections titels results, discussion, methods (<3000 words en sus). Less than 10 figures and tables.

\author{H. Graef} \affiliation{Laboratoire de Physique de l'Ecole Normale Sup\'erieure, Ecole normale sup\'erieure-PSL University, Sorbonne Universit\'e, Universit\'e Paris Diderot, Sorbonne Paris Cit\'e, CNRS, 24 rue Lhomond, 75005 Paris France}
\affiliation{CINTRA, UMI 3288, CNRS/NTU/Thales, Research Techno Plaza, 50 Nanyang Drive, 637553, Singapore}
\affiliation{School of Electrical and Electronic Engineering, Nanyang Technological University, 50 Nanyang Avenue, Singapore 639798, Singapore}
\author{Q. Wilmart}  \affiliation{Laboratoire de Physique de l'Ecole Normale Sup\'erieure, Ecole normale sup\'erieure-PSL University, Sorbonne Universit\'e, Universit\'e Paris Diderot, Sorbonne Paris Cit\'e, CNRS, 24 rue Lhomond, 75005 Paris France}
\author{M. Rosticher}  \affiliation{Laboratoire de Physique de l'Ecole Normale Sup\'erieure, Ecole normale sup\'erieure-PSL University, Sorbonne Universit\'e, Universit\'e Paris Diderot, Sorbonne Paris Cit\'e, CNRS, 24 rue Lhomond, 75005 Paris France}
\author{D. Mele}  \affiliation{Laboratoire de Physique de l'Ecole Normale Sup\'erieure, Ecole normale sup\'erieure-PSL University, Sorbonne Universit\'e, Universit\'e Paris Diderot, Sorbonne Paris Cit\'e, CNRS, 24 rue Lhomond, 75005 Paris France}
\author{L. Banszerus}
\affiliation{2nd Institute of Physics, RWTH Aachen University, 52074 Aachen, Germany}
\author{C. Stampfer}
\affiliation{2nd Institute of Physics, RWTH Aachen University, 52074 Aachen, Germany}
\author{T. Taniguchi}
\affiliation{Advanced Materials Laboratory, National Institute for Materials Science, Tsukuba,
Ibaraki 305-0047,  Japan}
\author{K. Watanabe}
\affiliation{Advanced Materials Laboratory, National Institute for Materials Science, Tsukuba,
Ibaraki 305-0047, Japan}
\author{J-M. Berroir} \affiliation{Laboratoire de Physique de l'Ecole Normale Sup\'erieure, Ecole normale sup\'erieure-PSL University, Sorbonne Universit\'e, Universit\'e Paris Diderot, Sorbonne Paris Cit\'e, CNRS, 24 rue Lhomond, 75005 Paris France}
\author{E. Bocquillon} \affiliation{Laboratoire de Physique de l'Ecole Normale Sup\'erieure, Ecole normale sup\'erieure-PSL University, Sorbonne Universit\'e, Universit\'e Paris Diderot, Sorbonne Paris Cit\'e, CNRS, 24 rue Lhomond, 75005 Paris France}
\author{G. F\`eve}
 \affiliation{Laboratoire de Physique de l'Ecole Normale sup\'erieure, Ecole normale sup\'erieure-PSL University, Sorbonne Universit\'e, Universit\'e Paris Diderot, Sorbonne Paris Cit\'e, CNRS, 24 rue Lhomond, 75005 Paris France}
\author{E.H.T. Teo}
\affiliation{School of Electrical and Electronic Engineering, Nanyang Technological University, 50 Nanyang Avenue, Singapore 639798, Singapore}
\affiliation{School of Materials Science and Engineering, Nanyang Technological University, 50 Nanyang Avenue, Singapore 639798, Singapore}
\author{B. Pla\c{c}ais} \email{bernard.placais@lpa.ens.fr}
 \affiliation{Laboratoire de Physique de l'Ecole Normale Sup\'erieure, Ecole normale sup\'erieure-PSL University, Sorbonne Universit\'e, Universit\'e Paris Diderot, Sorbonne Paris Cit\'e, CNRS, 24 rue Lhomond, 75005 Paris France}

\begin{abstract}
Dirac fermion optics  exploits the refraction of chiral fermions across optics-inspired Klein-tunneling barriers defined by high-transparency p-n junctions. We consider the corner reflector (CR) geometry introduced in optics or radars.  We fabricate Dirac fermion CRs using bottom-gate-defined barriers in hBN-encapsulated graphene. By suppressing transmission upon multiple internal reflections, CRs are sensitive to minute phonon scattering rates. We report on doping-independent CR transmission in quantitative agreement with a simple scattering model including thermal phonon scattering.  As a new signature of CRs, we observe Fabry-P\'erot oscillations at low temperature, consistent with single-path reflections. Finally, we demonstrate high-frequency operation which promotes CRs as fast phonon detectors. Our work establishes the relevance of Dirac fermion optics in graphene and opens a route for its implementation in topological Dirac matter.
\end{abstract}

\maketitle

 Since Landauer's work in the fifties [\onlinecite{Landauer1957ibm}], we know that electric transport can be described by the scattering of electronic waves, in close analogy with the transmission of light in matter. The electron-photon analogy is even more adequate in graphene due to the linear dispersion of massless Dirac fermions and their sublattice pseudospin polarization [\onlinecite{Allain2011epjb}]. For example the refraction of Dirac fermion at a p-n junction obeys electronic variants of Snell-Descartes and Fresnel relations [\onlinecite{Katnelson2006nphys,Cheianov2006prb,Cheianov2007Science,Cayssol2009prb}] with an  optical index proportional to the Fermi energy. Negative refraction and strong forward focusing effects could be measured  [\onlinecite{Lee2015nphys,Chen2016science}] using high-mobility hBN-encapsulated graphene [\onlinecite{Banszerus2015sciadv,Banszerus2016nl}]. Dirac Fermion optics (DFO) naturally explains the large transmission of a Klein-tunneling potential barrier  [\onlinecite{Huard2007prl,Shytov2008prl,Urban2011prb,Young2009nphys,Stander2009prl,Rickhaus2013ncom,Maurand2014carbon,Jung2016nl}], when regarded as an optical plate. The next step is to adapt the optics toolbox to Dirac fermions using  p-n junctions as high-transmission diopters and electrostatically-shaped Klein tunneling barriers to implement basic refracting functions like reflectors.

 Ballistic transistors exploiting total internal reflection across a prism-like, saw-tooth-shape, barrier have been proposed [\onlinecite{Jang2013pnas,Wilmart2014twoDm}] and demonstrated  [\onlinecite{Morikawa2017sst}] showing however rather limited on/off capabilities. Similar geometries, combining p-n junctions and graphene-edge reflections, have been proposed [\onlinecite{Sajjad2011apl,Sajjad2012PRB,Sajjad2013acsnano}] and measured [\onlinecite{Low2009prb,Williams2011nnano,Sutar2012nl}]. They  eventually suffer from spurious edge scattering  [\onlinecite{Walter2018prl,Elahi2018arxiv}]. Very recently, three-terminal ballistic switches have been demonstrated [\onlinecite{Wang2018arxiv,Elahi2018arxiv}], showing robust on/off ratios $\sim 5$. More involved DFO-based systems have also been considered, like Mie-scattering devices [\onlinecite{Caridad2016ncomm}], pinhole collimators [\onlinecite{Barnard2017ncomm,Liu2017prl}], Dirac fermion microscopes [\onlinecite{Bogglid2017ncomm}], and chaotic DFO systems [\onlinecite{Xu2018prl}]. Beyond graphene, the interest in DFO extends to black phosphorus [\onlinecite{Zhenglu2017nl}], borophene [\onlinecite{Zhang2018prb}], surface states of topological insulators [\onlinecite{Hassler2010prb}],  or massive states of topological matter [\onlinecite{Betancur2018jpcm,HungNguyen2018prb}].

  The first aim of the present work is to put DFO principles to quantitative test in the stringent geometry of a corner reflector using state-of-the-art bottom-gated hBN-encapsulated graphene. Right-angle prism CRs of Ref.[\onlinecite{Wilmart2014twoDm}] are two-dimensional (2D) Dirac fermion variants of  lunar laser retro reflectors [\onlinecite{Alley1965jgr}], or radar corner reflectors [\onlinecite{Garthwaite2013ieee}]. When compared to their optic counterpart, graphene CRs benefit from a gate-tunable refraction index ratio, in a range $n_r=-6\rightarrow 6$ exceeding that of  light refractors in the visible ($n_r\lesssim2.3$ in diamond) or infrared ($n_r\lesssim4$ in semiconductors) range.  Full suppression of electronic transmission relies on multiple ballistic total internal reflections predicted for $n_r\gtrsim2.5$ according to Ref.[\onlinecite{Wilmart2014twoDm}]. A second result of this study is that the residual CR transmission is ultimately limited by minute acoustic phonon scattering in the barrier, which acts as a cutoff time/length for these multiple internal reflections. This effect explains the modest on/off ratios of CR-transistors, which precludes their use for logic applications. When put in perspective with their excellent dynamical properties, it also promotes  CRs as valuable ballistic phonon detectors at low temperature.

\begin{figure}[ht]
	\centerline{\includegraphics[width=12cm]{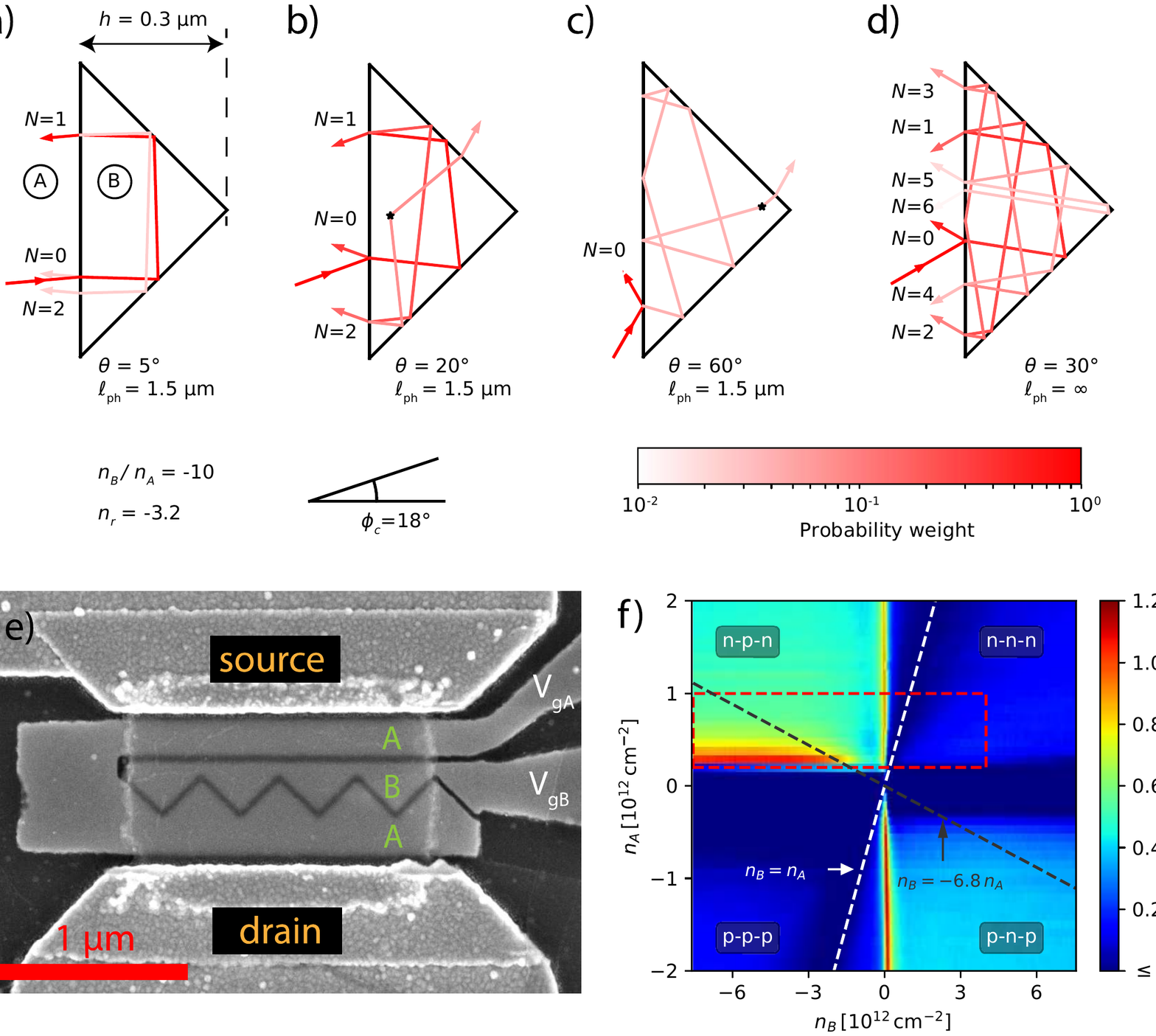}}
	\caption{\textbf{The reflector device.} (a-d) Calculated trajectories in a corner reflector with a critical angle $\phi_c=18^{\circ}$, corresponding to a doping ratio of 10 between the regions $A$ and $B$. At low incidence angle, the probability of the fermion to be reflected after one round trip is close to unity. When the incidence angle is slightly oblique, fermions have a high probability of staying trapped within the prism for multiple round-trips, and become susceptible to phonon scattering, which helps them escape the prism. At high incidence angle, the probability of being reflected before entering the prism is close to unity. (e) Annotated SEM micrograph of the device (sample CR-H9.4). Two bottom gates $V_{gA}$ and $V_{gB}$ allow for local control of the DFO refractive index in the access ($A$) and barrier ($B$) regions. hBN-encapsulated graphene is visible as a transparent rectangle, contacted at the edge by source and drain electrodes. (f) Color plot of the device resistance as a function of the charge carrier densities in regions $A$ and $B$. The shading of the four quadrants is typical for a double-junction: low resistance (dark blue) in the unipolar quadrants and higher resistance (light blue) in the bipolar quadrants. The resistance minimum $R(n_A=n_B)$ (contact resistance) was subtracted from the data. The white dashed line indicates $n_A=n_B$, consistent with the resistance minimum, whereas the black dashed line indicates $n_B=-6n_A$, the criterion for total internal reflection. In the following we will focus on the data within the red dashed box.}
	\label{DFO_principles.fig1}
\end{figure}

\section{Results}

As an introduction to reflector principles, figures \ref{DFO_principles.fig1}-(a-d)  show calculated electron trajectories (Supplementary material (SM) and below) for typical incidence angles $\theta$ in the presence of a finite phonon scattering length $\ell_{ph}=1.5\;\mathrm{\mu m}$ (see below). At low incidence ($\theta=5^\circ$ in Fig.\ref{DFO_principles.fig1}-a) quasi-total reflection is achieved upon a single dwell cycle. The dwell length $L_1=2h$, where $h=0.3\;\mathrm{\mu m}$ is the prism height, being smaller than $\ell_{ph}$, reflection is insensitive to phonon scattering. At larger incidence ($\theta=30^\circ$ in Fig.\ref{DFO_principles.fig1}-d) a similar reflection amplitude requires  $N\sim 5$  ballistic cycles, each cycle contributing to the total reflection as shown by the color code of rays in the figure. The $N$-path cycle length, $L_N\simeq NL_1$ in right angle CRs, is independent of the impinging position and eventually exceeds $\ell_{ph}$ leading to the breaking of  ballistic trajectories. The two central sketches, for $\theta=20^\circ$ in Fig.\ref{DFO_principles.fig1}-b and $\theta=60^\circ$ in Fig.\ref{DFO_principles.fig1}-c, illustrate this effect. Scattering induces a finite barrier leakage after a cycle number $N\sim l_{ph}/2h\simeq2.5$. As a matter of fact, scattering breaks the regularity of ballistic trajectories eventually leading electron rays to enter the otherwise forbidden p$^+$-n junction transmission window, for an incidence angle to the exit juncion $\phi\lesssim\phi_c\simeq 18^\circ$. Scattering can be due to interface roughness,  frozen disorder or, as in our experiment, to thermal phonons in which case a statistical averaging of scattering events becomes relevant. Note that due to pseudo-spin conservation, scattering in graphene has a drastic effect on electron trajectories, with a prominently $\pm\pi/2$ momentum rotation [\onlinecite{Monteverde2010prl}].\\

\textbf{Sample description}. Figure \ref{DFO_principles.fig1}-e is a scanning electron microscope (SEM) image of sample CR-H9.4. The fabrication is detailed in the Method section and SM. Eight similar devices have been fabricated and characterized, which are described in the SM. All samples are embedded in a three-port coplanar wave guide (inset of Fig.\ref{DFO_high-frequency.fig4}-a below) for DC and GHz characterization  which are performed in a variable temperature $40\;\mathrm{GHz}$ probe-station. One distinguishes in Fig.\ref{DFO_principles.fig1}-e  the semi-transparent  hBN-encapsulated graphene sample of dimensions $L\times W=0.8\times1.6\;\mathrm{\mu m}$. It has a bottom hBN thickness of $9\;\mathrm{nm}$ giving a gate capacitance  $C_g\approx 3\;\mathrm{mF/m^2}$. The doping window, $|n_{A,B}|=0.25\rightarrow 8\times10^{12}\;\mathrm{cm^{-2}}$, is limited by device integrity at high density and access resistance or spurious mirage effects due to charge puddles at low density. It corresponds to  electronic wave lengths $\lambda_{A,B}=\sqrt{4\pi /|n_{A,B}|}=15\rightarrow 70\;\mathrm{nm}$, a widely tuneable index ratio $n_r=\sqrt{|n_B/n_A|}=1\rightarrow6$, and n-p$^+$ junction critical angles $\arcsin{n_r^{-1}}= 10^\circ\rightarrow 90^\circ$.  Also seen in the Fig.\ref{DFO_principles.fig1}-e are the source and drain Cr/Au edge contacts, the refracting barrier gate (B) and the common source and drain access gate (A). Gate metallizations, of nominal thickness $25\pm5\;\mathrm{nm}$, are made of tungsten for DC devices and gold for GHz devices. A thin slit of width $25\;\mathrm{nm}$ is etched to secure inter-gate isolation and achieve steep p-n junctions. Together with the small bottom-hBN thickness it defines the junction length $d\approx 30\;\mathrm{nm}$ (verified by a COMSOL simulation) and provides high-transparency junctions of transmission ${\cal T}_{np^+}\sim0.5$. Our CR design fulfills geometrical optics conditions, $d\lesssim \lambda_{A,B}\ll h<\ell_{ph}$, where  $\ell_{ph}(T)=0.8\rightarrow24\;\mathrm{\mu m}$ in our working range $T=10\rightarrow300\;\mathrm{K}$  [\onlinecite{Graef2018jphysmat}]. The barrier consists of four connected right-angle prisms.  Their overlap secures a minimum gate length of $100\;\mathrm{nm}$ preventing direct source-drain tunneling. The short access length of $\sim 0.2\;\mathrm{\mu m}$, favors ballistic transport with a $\cos\theta$ distribution of the incidence angles. Owing to the n-doping of Cr/Au contacts, we work in the n-p$^+$-n regime where the highly doped barrier fully controls the transmission  ${\cal T}_{CR}(n_A,n_B)$ of the $M_A=k_AW/\pi$ access modes ($M_A=40\rightarrow80$  for $n_A=0.25\rightarrow 1\times\;10^{12}\;\mathrm{cm^{-2}}$).

Figure \ref{DFO_principles.fig1}-f is a color plot of the CR resistance in a broad range of access and barrier doping. The vertical Dirac-peak resistance line at $V_{gB}=0$ shows the independence of access and barrier doping control. The resistance vanishes at $n_A=n_B$. We have  subtracted a contact resistance $R_c\sim5\;\mathrm{k\Omega}$ that can be minimized down to $\sim 200\;\mathrm{\Omega}$ using a better technology [\onlinecite{Wang2018arxiv}] or sample design [\onlinecite{Yang2018nnano,Yang2018prl,Graef2018jphysmat}]. The CR effect shows up as a resistance resurgence for $|n_B|\gtrsim6 |n_A|$. Clearly visible in the n-p$^+$-n regime, it is elusive in the p-n$^+$-p regime, presumably due to spurious Klein-tunneling reflections at the n-doped contact. The red-dashed-rectangle in figure \ref{DFO_principles.fig1}-f delimits the working window analyzed below.\\

\begin{figure}[ht]
	\includegraphics[width=12cm]{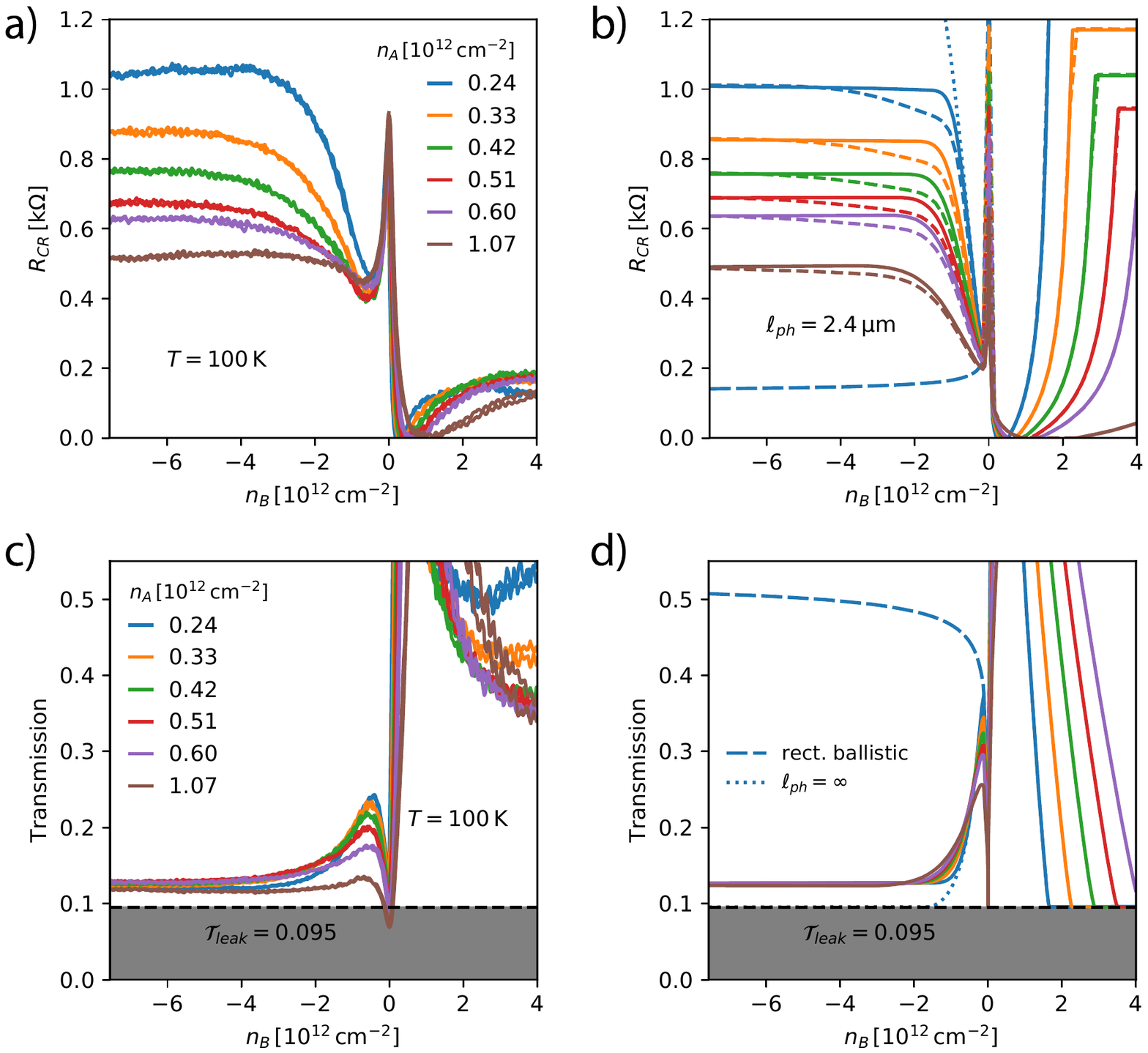}
	\caption{\textbf{Corner-reflector transfer characteristics in the incoherent regime ($T=100\;\mathrm{K}$): theory and experiment}.  (a,c) Experimental device resistance and transmission as a function of barrier doping $n_B$ for various access doping $n_A$ (c.f. red box in figure \ref{DFO_principles.fig1}-f). (b,d) Solid lines: transmission and resistance from simplified equation (SM-1), using a scattering length of $l_{ph}=2.4~\mathrm{\mu m}$ and a junction length of $d=30~\mathrm{nm}$, saturating in the n-p$^+$-n regime to the plateaus given by equation (\ref{Holger-formula}). Dashed lines: resistance from complete ray-tracing simulation. Dashed-dotted blue lines: resistance/transmission of a rectangular ballistic n-p-n barrier of same area and junction length and $n_A=0.24\times 10^{12}~\mathrm{cm^{-2}}$. Dotted blue lines: CR resistance/transmission with infinite scattering length in the same conditions. A leak transmission of 9.5\% was taken into account in all simulations. }
	\label{DFO_scattering.fig2}
\end{figure}

\textbf{The corner reflector transmission plateaus}. Figure \ref{DFO_scattering.fig2}-a shows a typical set of CR transfer characteristics $R_{CR}(n_A,n_B)$, taken at an intermediate temperature $T=100\;\mathrm{K}$ where $l_{ph}\sim2\;\mathrm{\mu m}$. Different curves correspond to different access doping $n_A>0$. The Dirac peak at barrier charge neutrality is accompanied by resistance resurgences for $|n_B|>n_A$ which are signatures of the CR effect. CR is fully developed in the bipolar regime, with resistance plateaus for $n_B\lesssim-6\;n_A\simeq -3\times10^{-12}\;\mathrm{cm^{-2}}$. The plateau resistance decreases for increasing access doping in accordance with the refraction principles prescribing a large index contrast. It eventually overshoots the Dirac point resistance, illustrating the efficiency of CRs in controlling the barrier transmission. Taking the $n_B=\pm4\times10^{12}\;\mathrm{cm^{-2}}$ values as reference  ``on'' and ``off'' states, we estimate an on-off ratio $\sim 5$ for the barrier resistance alone. The $n_A$-dependence is mostly determined by the number $M_A$ of access modes as seen in Fig.\ref{DFO_scattering.fig2}-c which shows a doping-independent transmission plateau,  ${\cal T}_{CR}=R_L/(R_{CR}+R_L)$ where $R_L=h/4M_Ae^2$ is the Landauer resistance. Note that these resistance plateaus, which are the signature of CRs, were not reached in the early investigation of Ref.[\onlinecite{Morikawa2017sst}] where $\sqrt{|n_B/n_A|}\lesssim 1.7$ remained below the CR regime conditions  $\sqrt{|n_B/n_A|}\gtrsim 2.5$.\\

\begin{figure}[ht]
	\centerline{\includegraphics[width=12cm]{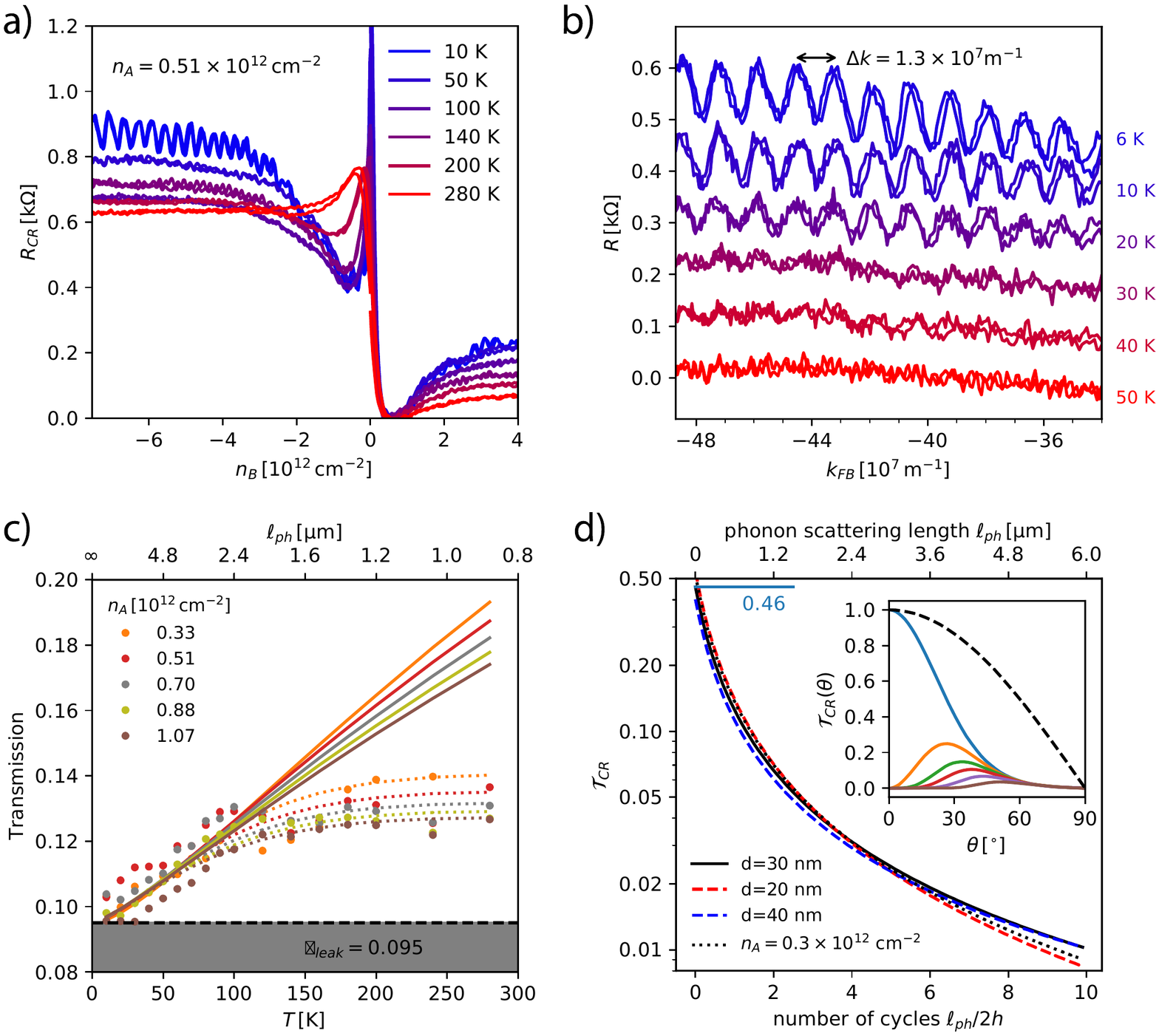}}
	\caption{\textbf{Temperature dependence of the corner-reflector transmission: coherence, phonon scattering and saturation.} (a) Device resistance at various temperatures $T=10\dots 280~\mathrm{K}$. The plateau resistance decreases as the temperature increases. (b) At low temperatures, Fabry-P\'erot type oscillations ($\Delta k = 1.3\times 10^7\;\mathrm{m^{-1}}$) are observed in the resistance, they disappear at $T\approx 40~\mathrm{K}$. (Resistance offset for clarity.) (c) Plateau transmission as a function of temperature for various access doping values $n_A$. Solid lines and dotted lines are calculated using Eq.(\ref{Holger-formula}) with $d=30\;\mathrm{nm}$ and $d=30\;\mathrm{nm} +2.25\;\mathrm{pm/K^2} \times T^2$ respectively.  (d) Plateau transmission Eq.(\ref{Holger-formula}) as a function of the number of round-trips in the CR, compared to the transmission of a single p-n junction ${\cal T}_{np^+} = 0.46$ (light blue bar). We used a junction length of $d=30~\mathrm{nm}$ and doping values $n_A=0.5$ and $n_B=-5\times 10^{12}~\mathrm{cm^{-2}}$ (black line). Varying the junction length (red and blue dashed lines) or the doping (black dotted line) only has a small effect on the overall shape of this curve. Inset: CR transmission as a function of incident angle, for various numbers of cycles. Blue line: transmission of a single n-p junction (zero cycles). Other color lines represent how the transmission is increasingly suppressed when increasing the number of round-trips (1, 2, 3, 5, 10). The transmission is weighted in Eq.(\ref{Holger-formula}) by the $\cos \theta$ factor (black dashed line).
	}
	\label{DFO_temperature.fig3}
\end{figure}

\begin{figure}[ht]
	\includegraphics[width=12cm]{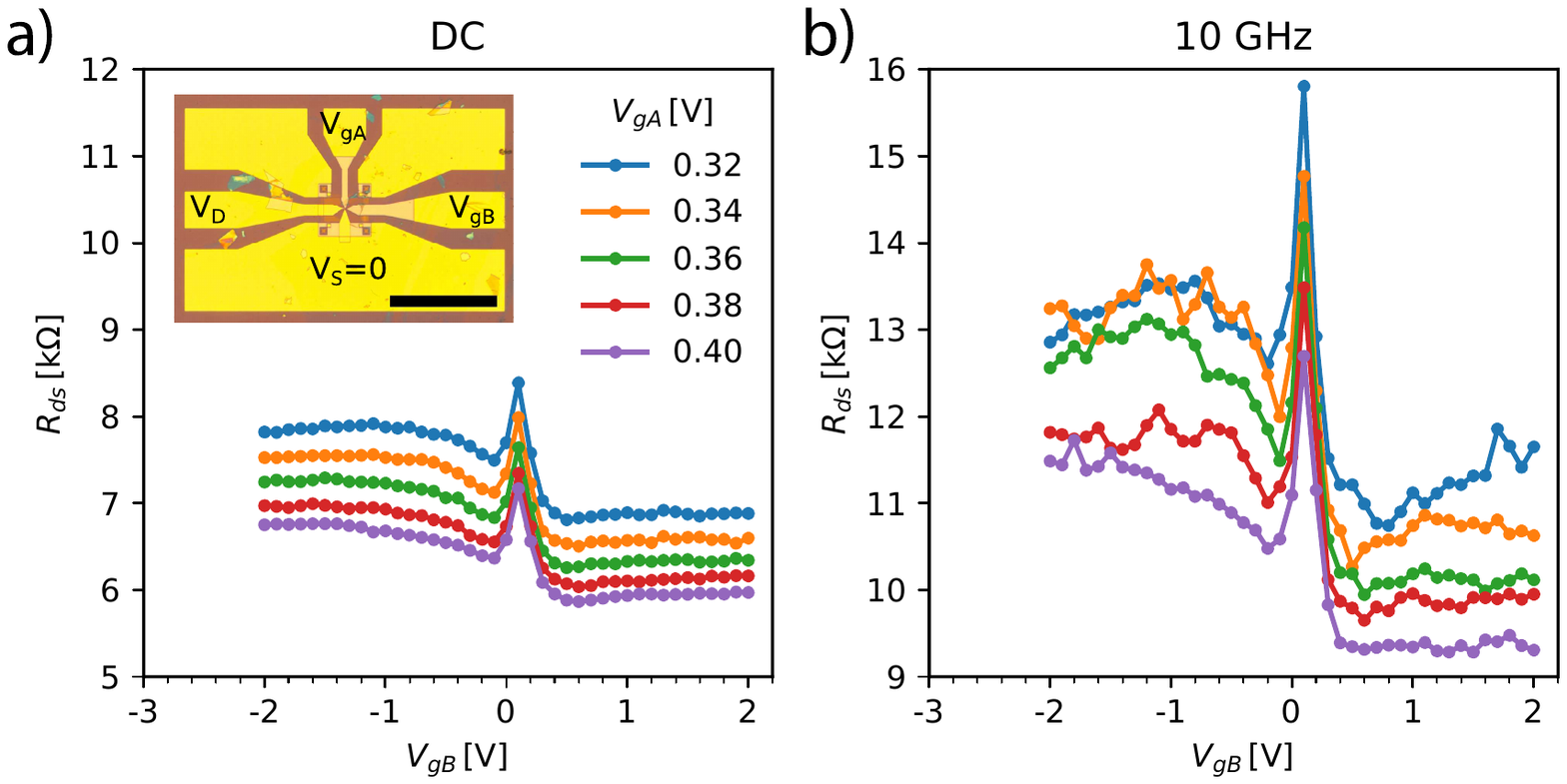}
	\caption{\textbf{High-frequency corner reflectors} (a) DC source-drain resistance $R_{ds}$ of sample CR-AuEG-17.ML (see SM) as a function of barrier gate voltage $V_{gB}$ for various access gate voltages. (b) High frequency source-drain resistance.}
	\label{DFO_high-frequency.fig4}
\end{figure}

\textbf{Phonon scattering limited corner reflector transmission}. The CR transmission can be calculated with semi-classical ray-tracing simulations [\onlinecite{Wilmart2014twoDm,Elahi2018arxiv}], taking into account the number of ballistic modes in the access region, the transmission ${\cal T}(\phi)$ of the three junctions as a function of refraction angle $\phi$ and multiple reflections inside the barrier. Accounting for residual quasi-elastic phonon scattering in this model amounts to randomizing the direction of propagation of the fermions after traveling an average length $\ell_{ph}$ (details in SM). In the special case of a right-angle prism, due to the existence of recurrent trajectories, an analytic formula for the prism transmission could be derived (see equation (SM-1)), which reduces to the following compact formula in the CR regime:

\begin{equation}
	\label{Holger-formula}
	{\cal T}_{CR}(T)\approx\int_{0}^{\frac{\pi}{2}}{\cal T}_{np}(\theta)\left[1-{\cal T}_{np}(\theta)\right]^{\ell_{ph}/2h}\cos\theta\; d\theta  \hspace{2cm} (\mathrm{for}\;|n_B| \gtrsim 6 |n_A|)\;,
\end{equation}
where the $\cos\theta$ factor accounts for a uniform distribution of the transverse momentum ($k_y = k_A \sin{\theta}$) in the access. A marked difference between Eq.(SM-1) and Eq.(\ref{Holger-formula}) is that the Snell-Descartes relation ($k_A \sin\theta = k_B \sin\phi$) no longer appears explicitly in the expression for the CR transmission, which becomes entirely determined by the junction transparency and mean free path.  Eq.(\ref{Holger-formula}) accounts for the statistical nature of phonon scattering by interpolating the integer number of cycles $N$ to continuous values  $\ell_{ph}/2h$. The $\ell_{ph}=0$ limit reduces to the bare n-p$^+$-junction transparency $\langle{\cal T}_{np}\rangle_\theta$.  Following Ref.[\onlinecite{Wilmart2014twoDm}], we use the Cayssol-Huard expression for the junction transmission [\onlinecite{Cayssol2009prb}] (electronic variant of the Fresnel-relations in optics):
\begin{equation}
	\label{Cayssol-Huard-transmission}
	{\cal T}_{np}(\theta)=1-\frac{\sinh{(\pi w \kappa^{+-})}\sinh{(\pi w \kappa^{-+})}}{\sinh{(\pi w \kappa^{++})}\sinh{(\pi w \kappa^{--})}}\; ,
\end{equation}
with  $\kappa^{\rho \sigma}=k_p(1+\rho \cos{\theta})-k_n(1-\sigma \cos{\phi})$ and $w=d/2\ln(10)$, where $d$ is the junction length [\onlinecite{Wilmart2014twoDm}]. The link between the incident angle $\theta$ and the transmitted angle $\phi$ is given by the equivalent of the Snell-Descartes law: $k_p\sin{\phi}=k_n\sin{\theta}$. Equation (\ref{Cayssol-Huard-transmission}) is suited for the asymmetric sharp junctions considered here. Alternatively, one can rely on the  Cheianov-Fal'ko's gaussian  expression for smooth symmetric junctions $T_{np}(\theta)\simeq \exp(-\pi k_{n} d\sin^2\theta)$ [\onlinecite{Cheianov2006prb}], suitably modified for asymmetric doping [\onlinecite{Sajjad2012PRB,Elahi2018arxiv}]. As shown in the inset of Fig.\ref{DFO_temperature.fig3}-d, ${\cal T}_{np^+}(\theta)$ vanishes for $\theta\gtrsim50^\circ$ in our sample. Eq.(\ref{Holger-formula}) is very useful for the design of Dirac fermion reflectors and the analysis of their properties. As a practical criterion we can estimate the number of cycles,  $N(\theta)\simeq(\log{\epsilon}-\log{{\cal T}_{np^+}(\theta)})/(1-{\cal T}_{np^+}(\theta))\sim5$, needed to achieve a $1-\epsilon=0.99$ reflection coefficient for ${\cal T}_{np^+}(\theta)\sim0.5$ ($\theta\simeq30^\circ$ in our device).

Figure \ref{DFO_scattering.fig2}-b shows the calculated CR resistance $R_{CR}(n_A,n_B)$ for $d=30\;\mathrm{nm}$, $h=0.3\;\mathrm{\mu m}$, and $\ell_{ph}=2.4\;\mathrm{\mu m}$ mimicking  $T=100\;\mathrm{K}$ conditions. A constant leakage transmission ${\cal T}_{leak}=0.095$ has been added to account for device imperfections. The analytic formula (SM-1) quantitatively reproduces the main features of experimental data in Fig.\ref{DFO_scattering.fig2}-a, including the plateau at negative $n_B$. The plateau is a mere consequence of the saturation of ${\cal T}_{np^+} (\theta)$ in increasing barrier doping $|n_B| \gg |n_A|$ according to Eq.(\ref{Cayssol-Huard-transmission}). In Eq.(SM-1) the model also predicts a strong resistance resurgence in the n-n$^+$-n regime which is not observed experimentally, a  discrepancy that remains to be clarified. Such a contrast between unipolar and bipolar regimes may stress the importance of the negative barrier index, or equivalently pseudo-spin conservation, in enforcing CR reflection conditions. The ray-tracing simulations are added as dashed lines in the figure. The overlap of the two types of simulations confirms the relevance of equations (SM-1) and (\ref{Holger-formula}) in describing right-angle prism CRs. Figure \ref{DFO_scattering.fig2}-d shows that the calculated transmission  ${\cal T}_{CR}(n_A,n_B)$ is indeed independent of doping in the CR regime in agreement with experiment.

Also shown in  Figs.\ref{DFO_scattering.fig2}-b and \ref{DFO_scattering.fig2}-d are the resistance/transmission of a conventional rectangular  Klein-tunneling (KT) barrier (blue dashed-dotted lines) for $n_A=0.24\times 10^{12}\;\mathrm{cm^{-2}}$. It is essentially controlled by the transmission of the n-p$^+$ barrier-entry junction, the strong forward focusing entailing a quasi total transmission at the parallel barrier-exit junction.  Inspection highlights the strong suppression of  transmission in CRs, well below the KT-barrier transmission value ${\cal T}_{KT}\simeq 0.4$. It also shows that the resistance dips at $n_B\simeq -n_A$ closely approach the rectangular barrier limit and can therefore be used to estimate the n-p$^+$-junction transparency $\langle{\cal T}_{np^+}\rangle_\theta$. The fully ballistic case of Ref.[\onlinecite{Wilmart2014twoDm}], ${\cal R}_{CR}(\ell_{ph}=\infty)$ and ${\cal T}_{CR}(\ell_{ph}=\infty)$  (blue dotted lines), is added for comparison. The parameter-free agreement between model and experiment concerns the existence of plateaus, their barrier doping threshold, as well as the access doping dependence. It constitutes a strong evidence that DFO principles can be implemented in a functional device on a quantitative basis. The value of the plateau transmission itself involves an extrinsic leakage ${\cal T}_{leak}$, which sets a limit to the on/off ratio of CRs at low temperatures. It can be minimized with a better sample design minimizing edge effects, with e.g. a larger number of prisms (and extension $W$) or an edge-free Corbino geometry.\\

\textbf{Temperature dependence of corner reflector transmission}. We now consider the effect of thermal electron and phonon excitations on the coherent and incoherent response in the CR transmission regime.    Figure \ref{DFO_temperature.fig3}-a is a plot of ${R}_{CR}(n_B,T)$ at $n_A=0.5\times 10^{12}\;\mathrm{cm^{-2}}$ in increasing temperatures. CR plateaus decrease with temperature implying a remarkable negative temperature coefficient of the resistance. At the lowest temperatures ($T=10\;\mathrm{K}$ in the figure),  Fabry-P\'erot oscillations superimpose on the plateaus, indicating constructive and destructive electronic interference. This distinctive property of CRs is elusive in obtuse-prism geometries [\onlinecite{Morikawa2017sst}], it reflects the constant single-path dwell length $L_1=2h$ of right angle prisms (Fig.\ref{DFO_principles.fig1}-a). FP oscillations, mostly visible in the n-p$^+$-n  regime, are $k_{p^+}$ periodic in  Fig.\ref{DFO_temperature.fig3}-b with a period $\Delta k=13\;\mathrm{\mu m^{-1}}$. Being independent of the access doping these oscillations can be unambiguously attributed to internal barrier reflections. They correspond to an optical length $2\pi/\Delta k\simeq 0.5\;\mathrm{\mu m}$ approaching the geometrical single cycle length $L_1=2h=0.6\;\mathrm{\mu m}$. They vanish above $30\;\mathrm{K}$ consistently with the expected thermal smearing temperature $T_{coh}=\pi\hbar v_F/(k_BL_1)\simeq 40\;\mathrm{K}$. The shape of the high-temperature characteristics, $T\gtrsim100\;\mathrm{K}$ in  Fig.\ref{DFO_temperature.fig3}-a, are qualitatively different: the Dirac peak and resistance dip are smeared and the CR merely behaves as a switch with a temperature-independent on/off ratio $\simeq 5$, reminiscent of that reported in the three-terminal ballistic switch of Ref.[\onlinecite{Wang2018arxiv}].

Figure \ref{DFO_temperature.fig3}-c gathers the measured plateau transmissions ${\cal T}_{CR}(n_A,T)$ (for $n_{B}<-5\times10^{12}\;\mathrm{cm^{-2}}$) over the full investigated access doping and temperature ranges. Solid lines are parameter-free theoretical predictions from Eq.(\ref{Holger-formula}) with ${\cal T}_{np^+}(n_B=\mathrm{const.})$ ($d=30\;\mathrm{nm}$) and $\ell_{ph}(T)=300/T\times 0.8\;\mathrm{\mu m}$ taken from Ref.[\onlinecite{Graef2018jphysmat}] measured in similar encapsulated samples. For simplicity we have neglected here deviations from the linear temperature dependence of the phonon scattering rates below the Bloch-Gr\"uneisen temperature [\onlinecite{Betz2012prl,Betz2013nphys}]. The good agreement of the $T$-dependence with our parameter-free model below $100\;\mathrm{K}$ supports the relevance of Eq.(\ref{Holger-formula}). Besides, we observe a saturation of transmission above $100\;\mathrm{K}$ at ${\cal T}_{CR}^{sat}\simeq 0.035$ for $\ell_{ph}(T)/2h\lesssim 4$, i.e.  well above the model application range $\ell_{ph}(T)/2h\gtrsim 1$. Such a transmission saturation cannot be explained by  a thermal activation process which would involve an excess transmission rather than a deficit. As a hint, we observe a concomitant thermal smearing of the $n_B=-n_A$ resistance dip for $T>100\;\mathrm{K}$ in Fig.\ref{DFO_temperature.fig3}-a, which means a decrease of the n-p$^+$-junction transparency and a thermal dilatation of the junction length $d(T)$. In this interpretation the saturation of transmission results from a compensation of the phonon-induced barrier transparency by a temperature-enhanced junction opacity.  As an illustration, we can reproduce in Fig.\ref{DFO_temperature.fig3}-c (dotted lines), experimental data with Eq.(\ref{Holger-formula}) taking an ad-hoc junction length  $d=30\;\mathrm{nm} +2.25\;\mathrm{pm/K^2} \times T^2$. In this analysis, the room-temperature junction length ($d\sim 200\;\mathrm{nm}$ at $T=280\;\mathrm{K}$) eventually approaches the length $h$ of the barrier itself which is qualitatively consistent with the observed smearing of Dirac point in Fig.\ref{DFO_temperature.fig3}-a. A proper account of this effect is beyond the validity domain of Eq.(\ref{Holger-formula}) and the scope of our paper which mostly focuses on phonon-scattering effects.

The good overall agreement between experiment and scattering theory in Figs.\ref{DFO_scattering.fig2} and \ref{DFO_temperature.fig3} consolidates the relevance of Eqs.(\ref{Holger-formula}) and (\ref{Cayssol-Huard-transmission}) in describing refraction properties of Dirac fermions. In particular it rules out the importance of junction roughness or frozen disorder in controlling CR's transmission. The former should lead to appreciable deviations from Fresnel relations  and the latter to a saturation of the phonon mean-free-path effect at low temperature, both of which are not observed. The effect of interface disorder is indeed  minimized here by a careful back-gate nano-patterning and its residual impact is reduced by a smooth junction  effect with $d\sim\lambda_{p^+}$. The effect of frozen disorder is minimized in our high-mobility encapsulated graphene samples.

The DFO principles at work in CRs are summarized in Fig.\ref{DFO_temperature.fig3}-d which shows the calculated plateau transmission ${\cal T}_{CR}(\ell_{ph}/2h)$ of Eq.(\ref{Holger-formula}) (solid black line). Additional lines have been drawn for nearby device parameters including the junction length; they illustrate the robustness of the CR effect, and its sensitivity as a \emph{ballistic-length meter}  to multi-cycle internal reflections. The inset shows ${\cal T}_{CR}(N,\theta)$ in increasing internal cycle number $N=0$--$10$. The $N=0$ case is the bare n-p$^+$ junction transmission collimating incident electrons within a $\pm 50^\circ$ aperture angle. It yields an angle-average transmission ${\cal T}_{np^+}\simeq0.46$ (blue bar in the main panel). Already prominent at $N=1$ for $\theta\lesssim 10^\circ$, the suppression extends to a wider angular domain in increasing $N$ reducing to a residual leakage ${\cal T}_{CR}\lesssim0.01$  at about  $\theta\sim45^\circ$ for $N=10$. \\

\textbf{Dynamical properties of corner reflectors}. Owing to the large ratio, $v_F/s\simeq 50$, of Fermi and sound velocities in graphene, corner reflectors are expected to have a picosecond time response, with a transit frequency $f_T\simeq v_F/\ell_{ph}\gtrsim100\;\mathrm{GHz}$ and a phonon time-of-flight $\tau\simeq h/s\simeq15\;\mathrm{ps}$. In order to assess this high-speed property we have fabricated a few CRs equipped with low-resistance gold gates suitable for GHz operation (see SM).  Figure \ref{DFO_high-frequency.fig4} compares the DC and $10\;\mathrm{GHz}$ resistance $R_{ds}$ at $T=60\;\mathrm{K}$ showing that CR properties are essentially preserved at $10\;\mathrm{GHz}$. This observation is consistent with the above estimate and theoretical predictions in Refs.[\onlinecite{Tan2017scirep,Pandey2018sst}]; it shows that CRs are as fast as conventional graphene transistors (see e.g. [\onlinecite{Pallecchi2011apl,Wilmart2016scirep}]). The combined high speed and phonon-scattering sensitivity of CRs offer promising perspectives in terms of ballistic phonon sensing. As a matter of fact, the phonon ballistic length, which is limited to a few hundreds of nanometers at room temperature, can exceed the CR barrier length at low temperature due to decreasing phonon-electron and phonon-phonon scattering rates [\onlinecite{Nika2009prb}]. Phonon time of flight measurements can thus be envisioned by merely monitoring transient CR resistance dips following the propagation of a remote phonon wave created by e.g. an optical pulse.

\section{Conclusion}

We have demonstrated  Dirac fermion corner reflectors using  state-of-the-art bottom-gate defined high-mobility graphene nano-transistors. We have characterized the CR transmission over a broad range of access/barrier doping, temperatures and frequencies. We have unveiled the existence of CR transmission plateaus and their dependence on the phonon scattering length in quantitative agreement with a simple scattering theory. Our modeling is further supported by the observation of Fabry-P\'erot oscillations in the quantum coherent regime which provides an independent measurement of the single path dwell length that is consistent with the multi-path length measured by the phonon scattering effect. We report on the saturation of CR transmission at high temperature attributed to a thermally induced junction opacity. Our experiment explains the longstanding issue of the finite on/off switching capabilities of reflectors which is limited at low temperature by device imperfections and at high temperature by the finite phonon scattering rate. We propose a new application of reflectors as ballistic phonon sensors, relying on their high speed and phonon-scattering sensitivity  below $100\;\mathrm{K}$. CRs enrich the family of Klein-tunneling based devices, including the recently demonstrated ballistic switches [\onlinecite{Wang2018arxiv}], contact barrier transistors [\onlinecite{Wilmart2016scirep}], and Zener-Klein transistors [\onlinecite{Yang2018nnano}]. Besides, the quantitative approach of DFO demonstrated in this work motivates new basic-physics studies in other Dirac matter systems such as topological materials or eventually bilayer graphene where refraction laws are modified due to anti-Klein tunneling effects [\onlinecite{Katnelson2006nphys}].

\section{Methods}

\textbf{Corner reflector fabrication techniques}. For the fabrication of nano-structured bottom gate electrodes, thin ($\sim$30 nm) films of tungsten (W) or gold (Au) were deposited using sputtering or evaporation on a high-resistivity Si/SiO$_2$ substrate. The patterning was done using two separate electron beam lithography (EBL) steps with positive poly(methyl methacrylate) (PMMA) resist for the $25\;\mathrm{nm}$ sawtooth line (defining the gap between the two gates) and the rough structures, each followed by reactive ion etching (SF$_6$ plasma for W [\onlinecite{Wilmart2016scirep}], Ar plasma for Au). In order to contact the gate electrodes with our GHz probe tips, we deposited a thicker ($150\;\mathrm{nm}$) layer of Au all over the coplanar waveguide (CPW) leading to the sample center (see inset figure \ref{DFO_high-frequency.fig4}a). Prior to subsequent fabrication steps, the gate electrodes were evaluated using  SEM and tested for short circuits.

High-mobility hBN-encapsulated graphene was fabricated according to reference \cite{Banszerus2015sciadv} and subsequently transferred using PPC-scotch tape-PDMS stamps \cite{Wang2013sci} on W electrodes and PMMA-poly(vinyl alcohol) stamps \cite{Banszerus2015sciadv} on Au electrodes. The stacks were then etched through an EBL-defined PMMA mask into a nearly rectangular shape covering only the active area (about $1.6\times1.3\;\mathrm{\mu m}$, see figure \ref{DFO_principles.fig1}a) of the device using a CHF$_3$/O$_2$ plasma. Finally, Cr/Au (5/100 nm) drain and source contacts were deposited after a last EBL step.\\

\textbf{DC and RF characterization}. The device characterization was carried out in a Janis cryogenic probe station operating between 6 and 300 Kelvin. The gate voltages were controlled using iTest Bilt voltage sources. The ``DC'' resistance of the device was measured using a Zurich Instruments HF2LI lock-in amplifier at 10.013 kHz with 5 mV amplitude in a voltage divider configuration.

The GHz frequency characterization was done using an Anritsu MS4644B vector network analyzer (VNA). In this case, the DC resistance was monitored simultaneously using another Bilt voltage source/meter, where the DC and GHz response of the device were decoupled from each other using bias tees. A standard short-open-load-reciprocal (SOLR) protocol was employed to calibrate the GHz wave propagation until the probe tips and the admittance parameters were de-embedded from parasitic capacitance by measuring a reference sample without graphene.

\begin{acknowledgments}
The research leading to these results have received partial funding from the the European Union ``Horizon 2020'' research and innovation programme under grant agreement No. 785219 ``Graphene Core'', and from the ANR-14-CE08-018-05 ``GoBN''. Authors wish to thank C. Voisin and S. Gigan for a critical reading of the manuscript before submission.
\end{acknowledgments}

\textbf{Author contributions} HG, QW and BP conceived the experiment. QW developed the bottom gate fabrication process. LB, CS, TT, KW, HG and MR participated to sample fabrication. HG conducted the measurements. HG, DM, JMB, GF, EB, EHTT and BP participated to the data analysis. QW and HG developed the ray-tracing simulations. HG and BP worked out the analytic model. HG and BP wrote the manuscript with contributions from the coauthors.

\textbf{Additional information} Competing financial interests: The authors declare no competing financial interests.

\end{document}